\documentclass[aip,apl,reprint,superscriptaddress]{revtex4-1}

\usepackage{graphicx}
\usepackage{color}
\usepackage{epstopdf}
\begin{document}

\title{Inscription of 3D waveguides in diamond using an ultrafast laser}

\author{Arnaud Courvoisier}
\affiliation{Department of Engineering Science, University of Oxford, Parks Road, Oxford, OX1 3PJ, UK}
\author{Martin J. Booth}
\affiliation{Department of Engineering Science, University of Oxford, Parks Road, Oxford, OX1 3PJ, UK}
\author{Patrick S. Salter}
\affiliation{Department of Engineering Science, University of Oxford, Parks Road, Oxford, OX1 3PJ, UK}
\email{patrick.salter@eng.ox.ac.uk}
\date{\today}

\begin{abstract}
Three dimensional waveguides within the bulk of diamond are manufactured using ultrafast laser fabrication. High intensities within the focal volume of the laser cause breakdown of the diamond into a graphitic phase leading to a stress induced refractive index change in neighboring regions. Type II waveguiding is thus enabled between two adjacent graphitic tracks, but supporting just a single polarization state. We show that adaptive aberration correction during the laser processing allows the controlled fabrication of more complex structures beneath the surface of the diamond which can be used for  3D waveguide splitters and Type III waveguides which support both polarizations. 
\end{abstract}


\maketitle

Diamond has long found use as a material with extreme mechanical properties, and is now rapidly gaining interest for photonic applications~\cite{Aharonovich2011}. High transmission over a very large spectral window and a high refractive index are attractive for light manipulation. Diamond also acts as a stable basis for a host of color centers which are promising for quantum enhanced technologies~\cite{Jelezko2006, Awschalom2007}. The nitrogen vacancy (NV) center is proving particularly useful not just for quantum processing~\cite{Bernien2013} but, also for a range of sensors with extreme sensitivity to magnetic field and temperature~\cite{Balsub2008, Maze2008, Neumann2013}. The strong Raman coefficient is effective for Raman lasers generating low cost lasing solutions at unusual wavelengths~\cite{Mildren2009, Kemp2015}. The high Raman coefficient is also useful for implementing solid state quantum memories, a vital component for any quantum optical technology enabling the storage and retrieval of quantum information~\cite{eilon2016}. In addition, diamond serves as an ideal platform for biophotonics due to a high level of biocompatibility~\cite{Picollo2013}. The efficiency of many of these technologies would be improved by the ability to confine and route light using a network of optical waveguides~\cite{clevenson2015}.

Previous demonstrations of waveguiding within diamond have been based upon the principle of selectively removing regions of diamond to generate an air-diamond interface. Surface RIB waveguides have previously been fabricated  using reactive ion etching (RIE) with photolithographic patterning~\cite{Prawer2008, Erdan2011}. Elegant nanobeam waveguides have recently been demonstrated through either angled~\cite{Burek2014} or undercut RIE etching~\cite{barclay2015}. Alternatively, direct ion microbeam writing has been used to create shallow subsurface multimode waveguides~\cite{Lagomarsino2010}. All of these approaches are constrained to the diamond surface, require extensive material processing to the potential detriment of the diamond and are difficult to interface with optical fibers. However, a different method for generating guiding structures in crystals has emerged over the past decade, whereby an ultrashort pulsed laser is used to modify the material~\cite{Chen2014}. In the majority of implementations, light is confined in regions of stress neighboring laser induced damage tracks, which is designated as waveguide based upon a Type II modification~\cite{Gross2015}. Previously this has been successfully applied to a range of crystals such as lithium niobate~\cite{osellame07b, thomas2011}, KDP~\cite{huang2015}, KTP~\cite{campbell2007b} and Ti:Sapphire~\cite{stoian2012}, whilst preserving the natural material properties. Here we extend the technique to fabricate three dimensional waveguides embedded in the diamond bulk.

Ultrashort pulsed laser fabrication inside diamond has been gaining interest due to possibilities related to the generation of 3D electrical circuitry. At the laser focus, the electric field is high enough for non-linear absorption leading to optical breakdown of the diamond lattice~\cite{Lagomarsino2016}. The structural modification generated, as identified through Raman microspectroscopy is revealed as graphitic and amorphous \textit{sp$^2$} bonded carbon intermixed with \textit{sp$^3$} bonded diamond~\cite{Kononenko2008, SalterSPIE2014}, while it has recently been shown that, in addition, the NV concentration may be enhanced in nearby regions~\cite{neuenshwander2016}. When the laser focus is traced through the diamond, a continuous damage track is created which is electrically conductive~\cite{Lagomarsino2013a, Sun2014b}, and has been successfully used for a range of radiation detectors~\cite{Oh2013b, Caylar2013a, Kononenko2013e}. Transmission electron microscopy and electron energy loss spectroscopy has been used to study thin slices of the laser-modified area, showing the presence of sub-micrometer patches of \textit{sp$^2$} bonded amorphous carbon, in addition to multiple dislocations of the diamond lattice~\cite{Unpub_FIB}. Indeed, due to the lower density of the \textit{sp$^2$} phase, compared to that of \textit{sp$^3$}, a strong localized stress field is generated within the surrounding pristine diamond, which we show here can act as an optical waveguide. Drawing on previous work demonstrating that a combination of high NA focusing and adaptive optics aberration correction is essential for accurate fabrication of graphitic tracks in diamond~\cite{Sun2014b}, we are able to generate Type III depressed cladding waveguides supporting both polarization states and 3D networks of Type II waveguides.

\begin{figure}[h!]
    \vspace{0cm}
    \begin{center}
        \includegraphics[width=8cm, natwidth=1660, natheight=482]{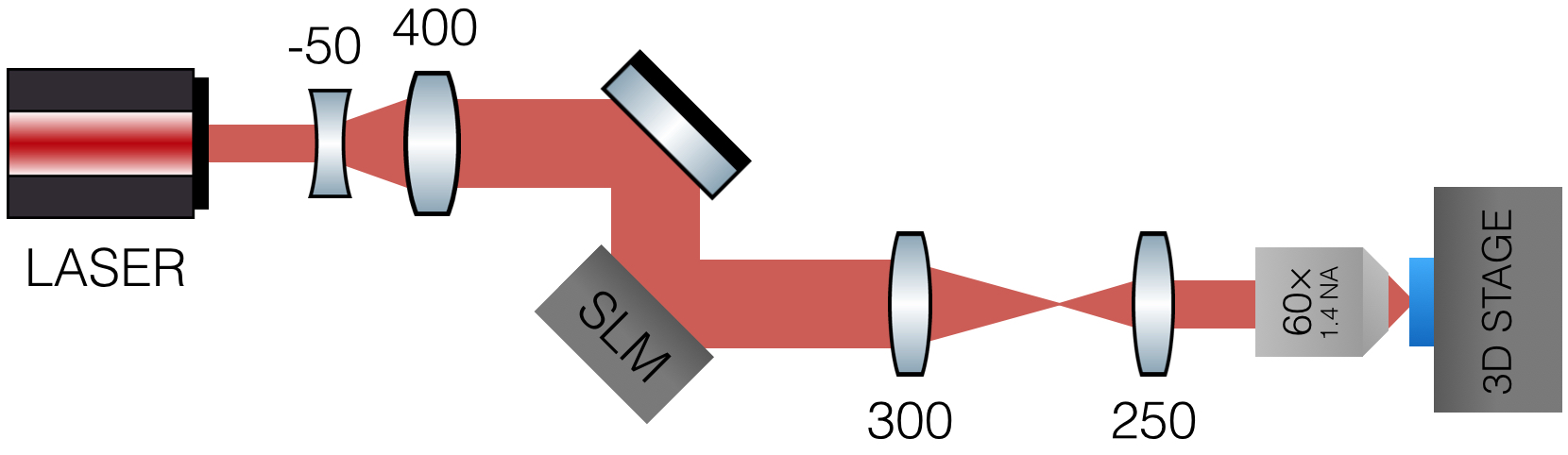} \\
        \vspace{-0.3cm}
        \caption[example]{\label{Figure0}\small{\emph{Simplified schematic of the ultrafast laser writing system. Focal lengths are displayed in millimeters beside each lens. SLM: Spatial Light Modulator.}}}
    \end{center}
    \vspace{-0.5cm}
\end{figure}

A schematic for the adaptive optics enabled laser fabrication system is displayed in Figure~\ref{Figure0} and described in detail elsewhere~\cite{huang2016}. In short,  a regeneratively amplified Ti:Sapphire laser (pulse duration 100 fs, pulse repetition rate of 1 kHz and central wavelength of 790 nm) was  expanded and directed onto a reflective liquid-crystal phase-only Spatial Light Modulator (SLM). The SLM was imaged onto the pupil plane of a high numerical aperture (NA) microscope objective (Olympus PlanApo 60$\times$ 1.4NA) via a 4$f$ system. The beam was thus focused into the diamond sample, which was mounted on a three-axis air bearing translation stage. The sample consisted of a 3~mm$\times$3~mm$\times$0.5~mm CVD single crystal diamond (Element 6) with all facets polished. 

The SLM was set to display a phase pattern to compensate the spherical aberration introduced by focusing into the diamond sample. The strong refractive index mismatch between diamond (n=2.4) and the objective immersion medium (n=1.52) introduced a depth-dependent spherical aberration~\cite{Salter2011}. Correction of this aberration enabled focusing at high numerical aperture and gave highly repeatable structural modifications with sub-micrometer dimensions\cite{Sun2014b} at a range of depths within the diamond. This allows accurate control over the position and shape of the focus for the manufacture of low-loss customized waveguides. 

\begin{figure}[h!]
    \vspace{0cm}
    \begin{center}
        \includegraphics[width=8cm, natwidth=1743, natheight=649]{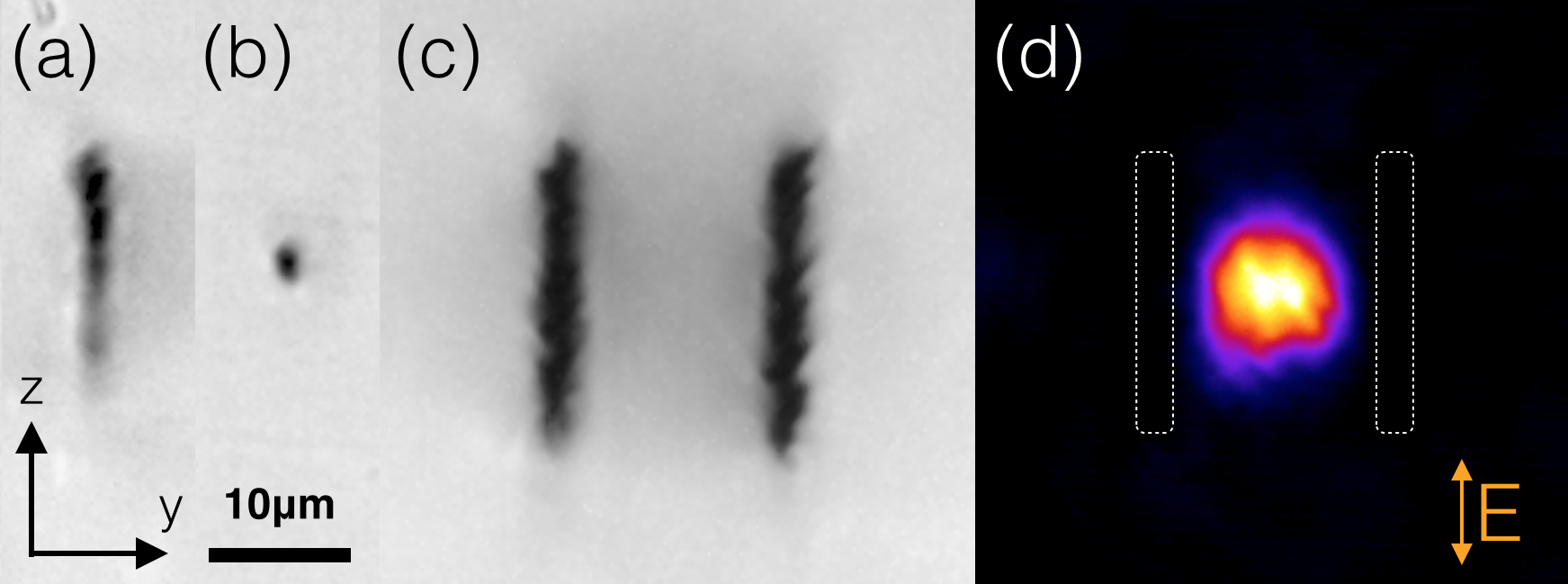} \\[0cm]
        \vspace{-0.3cm}
        \caption[example]{\label{Figure1}\small{\emph{Transmission microscope image of a single laser written graphitic track (a) without  and (b) with aberration correction, as seen from the sample's end-facet. (c) End-facet of the Type II double-line structure and (d) the near-field mode obtained for vertical (V) polarization.}}}
    \end{center}
    \vspace{-0.5cm}
\end{figure}

The diamond sample was translated through the laser focus perpendicular to the optic axis at a speed of 0.1~mm/s.  Without any aberration correction applied, the modification threshold pulse energy was 110~nJ at a depth of 50~$\mu$m beneath the diamond surface. The resulting  structures had a cross section which was significantly elongated axially (Fig~\ref{Figure1}(a)) and non-uniform along the length of the track. Using adaptive optics to compensate the aberration, the fabrication pulse energy could be reduced to below 30~nJ, generating highly uniform and repeatable graphitic tracks  with transverse and axial dimensions of 1~$\mu$m and 2$\mu$m respectively, as seen in Fig.~\ref{Figure1}(b). Due to the higher uniformity in the fabrication,  waveguiding structures were subsequently generated always using aberration correction during the processing. 

Type II waveguide structures were fabricated by laser writing pairs of graphitic tracks as shown in Fig.~\ref{Figure1}(c). The optimum transverse separation of the two damage lines was determined to be 17~$\mu$m for maximum transmission.  Each vertical graphitic line was fabricated using an axial multiscan approach, with 6 consecutive scans axially separated by 3~$\mu$m. We note that the axial separation of each scan is actually greater than the size of an individual graphitic track. However, this does not cause a discontinuous structure because the bottom graphitic layer is strongly absorbing and acts as a seed, with each subsequent graphitic track fabricated above being extended axially. 

The waveguides were characterized with light of wavelength 780~nm delivered via polarization maintaining fiber (PMF) butt coupled to the input facet of the sample without index matching gel. To observe the near field mode, the output facet of the sample was imaged onto a CCD camera by means of a microscope objective (Olympus ULWD MSPLAN 80$\times$, 0.75 NA). For loss measurements, the CCD was replaced by a power meter and an iris used to block unguided light. In contrast to laser written waveguides in other materials, there was no need to repolish the facet of the diamond following fabrication. Despite the strong aberration induced by the edge of the sample~\cite{Salter2012b}, fabrication is possible all the way up to the edge of the diamond, since the graphitic phase formed at the laser focus was highly absorbing and acted as a seed for further transverse fabrication.

The type II waveguides were found to be highly polarization dependent, analogous to studies of type II structures in other crystals~\cite{Burghoff2007, Chen2014}. There was strong guiding for vertically (V) polarized light (defined as along the $z$ axis in Fig.~\ref{Figure1}) while there was negligible transmission for the horizontal (H) polarization state. Considering just the V polarization, the overall insertion loss for the device was 4~dB. Taking into account the Fresnel reflection at the air-diamond interface as contributing a loss of 17\% at each facet, but neglecting any loss due to mode mismatch between the waveguide and input fiber, the propagation loss in the sample is estimated as 7.9~dB/cm. The waveguide exhibits single mode behavior in the horizontal direction (Fig.~\ref{Figure1}(d)) but it is possible to excite multiple modes when scanning the input fiber in the vertical direction. 

\begin{figure}[h!]
    \vspace{0cm}
    \begin{center}
        \includegraphics[width=8cm, natwidth=1606, natheight=1250]{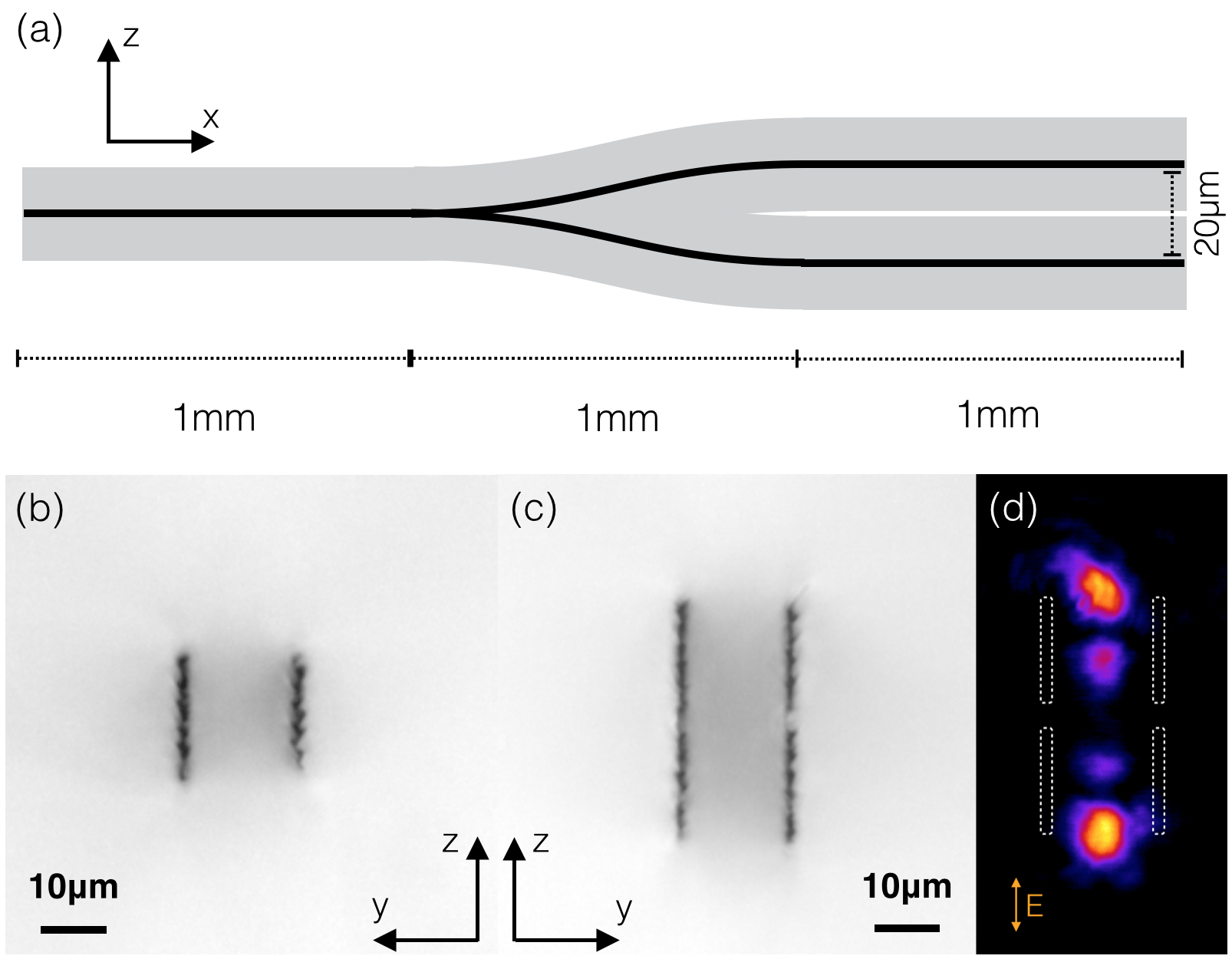} \\[0cm]
        \vspace{-0.3cm}
        \caption[example]{\label{Figure2}\small{\emph{(a) Schematic  structure of the vertical Y-splitter. Images from a transmission microscope showing the input (b) and output (c) facet of the splitter. (d) An image of the output near-field mode with a 50:50 power splitting.}}}
    \end{center}
    \vspace{-0.5cm}
\end{figure}

Using adaptive optics aberration correction, it is possible to fabricate identical structures at any depth within the diamond. Thus we were able to fabricate a vertical Y splitter from the Type II waveguides. The schematic structure is shown in Figure~\ref{Figure2}(a). The input facet comprises a single Type II waveguide, which splits over a propagation  distance of 1~mm into two identical waveguides axially separated by 20~$\mu$m at the output facet. During fabrication, the phase pattern displayed on the SLM was updated in closed loop using position feedback from the translation stages to ensure the aberration correction was always optimum. The coupling ratio between the two outputs can be varied by translating the input fiber along the z-axis. Thus it is possible to obtain a splitting ratio of 50:50 as seen in the near field mode image of Fig.~\ref{Figure2}(d). With a 50:50 splitting, the overall transmission of the structure is -7~dB, corresponding to a propagation loss of 18~dB/cm taking Fresnel reflections into account.  

\begin{figure}[h!]
    \vspace{0cm}
    \begin{center}
        \includegraphics[width=8cm, natwidth=1479, natheight=981]{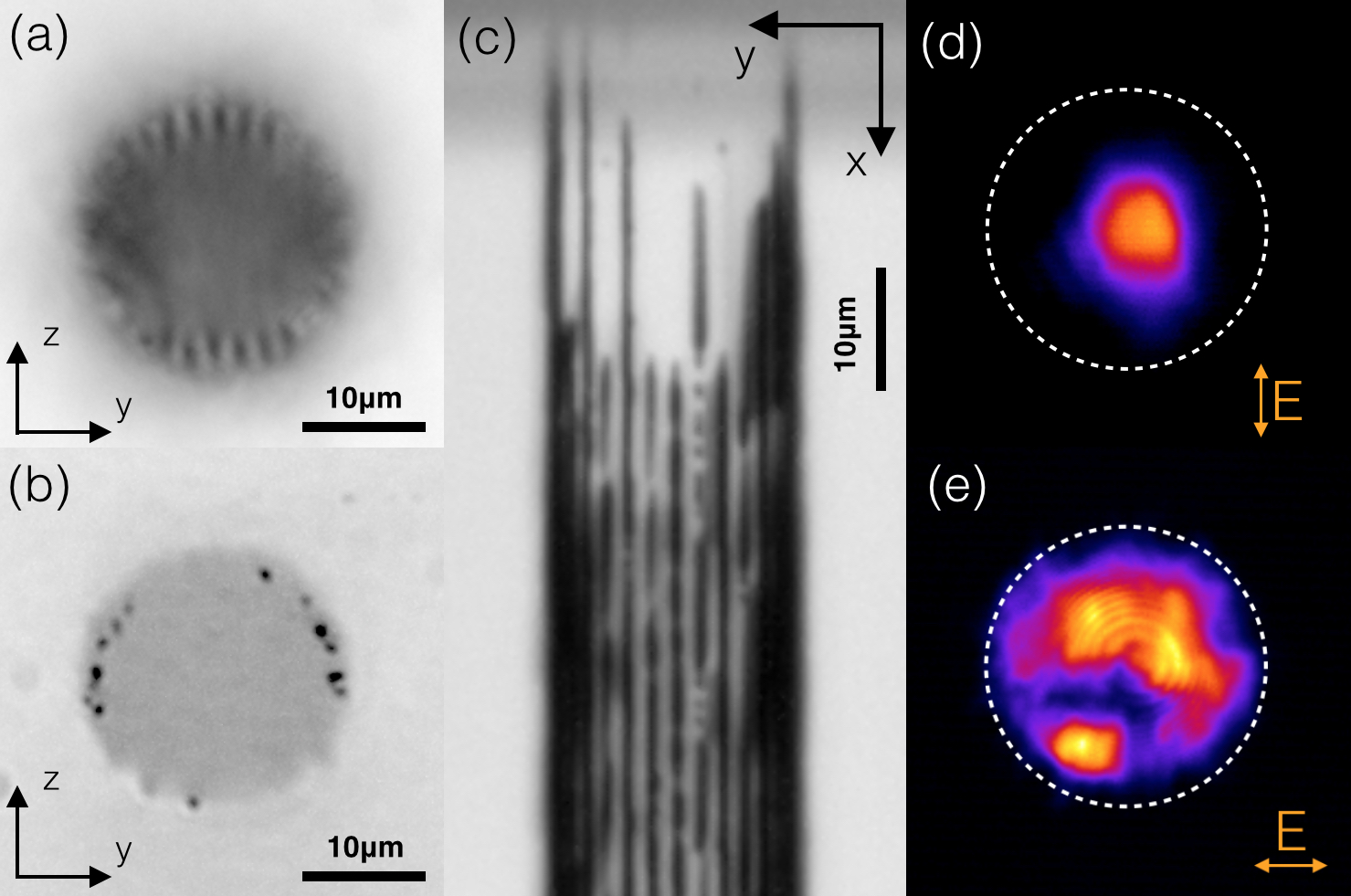} \\[0cm]
        \vspace{-0.3cm}
        \caption[example]{\label{Figure3}\small{\emph{Microscope images of the end-facet of the type III waveguide taken in (a) transmission focused slightly beneath the surface and (b) reflection focused on the sample surface. (c) Top-down transmission microscope image of the graphitic tracks comprising the waveguide. The edge of the sample is at the top of the image. Sample images of the near-field mode for V (d)  and H (e) polarizations.}}}
    \end{center}
    \vspace{-0.5cm}
\end{figure}

In order to fabricate Type III waveguides, which support both polarization states, 10~$\mu$m radius tubular structures  were generated as shown in Figure~\ref{Figure3}. 32 laser written graphitic tracks arranged in a circular manner were fabricated with a center-to-center spacing of 2~$\mu$m between each track.  The waveguides had insertion loss of -5.5~dB for the V mode and -11.5~dB for the H mode. Type III waveguides were multimode for both polarizations, with minimum loss found for light in the fundamental mode, as seen in Fig.~\ref{Figure3}~(d). We note that it would not be possible to fabricate such Type III structures without the use of adaptive optics during fabrication, as can be seen from comparison of the single graphitic tracks made without and with aberration correction in Fig.~\ref{Figure1}~(a) and (b).

The design of the type III structure is symmetric so is expected to be polarization insensitive. The difference in transmission for H and V polarizations  found in experiment is attributed to the fact that the laser focus is in fact asymmetric, as is each individual graphitic track (dimensions $\sim 1\times 2~\mu$m). Thus, the sides of the tube are continuous while top and bottom still have regions of unmodified diamond separating graphitic tracks, as can be seen in Figs.~\ref{Figure3}(a) and (c). The tubular structure is therefore expected to display an asymmetric stress field leading to the polarization dependent transmission. In principle, this can be mitigated by adjusting the relative positioning of the graphitic tracks that make up the tube, but in practice this becomes difficult as the accumulated stress can lead to cracking of the diamond. Indeed to avoid cracking of the diamond during manufacture of the Type III waveguides, it was necessary to reduce the fabrication pulse energy to 8~nJ. Such low pulse energies are at the fabrication threshold of the diamond, and as a result the graphitic tracks at the top and bottom of the  structure are discontinuous, as seen in Fig.~\ref{Figure3}~(c). The side walls remain continuous due to the axial overlap of the tracks which further contributes to the polarization dependent transmission. The reduced pulse energy additionally made it difficult to fabricate these Type III waveguides right to the edge of the diamond, as seen by the reflection microscope image of the end facet in Fig.~\ref{Figure3}~(b). Therefore a slightly higher coupling loss can be expected in this case.

In conclusion, we have demonstrated  inscription of 3D waveguides in the diamond bulk using an ultrafast laser, with examples of 3D waveguide splitters and Type III structures capable of guiding both polarization states. The ability to fabricate in 3D deep within the bulk and interface with optical fiber represents significant advantages for a range of integrated optics. However, values for the propagation loss are currently relatively high in comparison to those that can be achieved for surface nanobeam waveguiding structures~\cite{Burek2014}.  The propagation loss can be attributed to scattering and absorption from the \textit{sp$^2$} bonded carbon within the laser written tracks. It has been shown that fabrication of diamond using heavy ion implantation can lead to structural modifications with a smooth modulation of the real part of the refractive index~\cite{Lagomarsino2010, draganski2011}. In this study, we were unable to induce such Type I modifications, but utilizing different pulse repetition rates of the laser might potentially give access to different fabrication regimes where this becomes possible. It is apparent though that the incorporation of adaptive optics for aberration correction into the laser fabrication system is crucial for the controlled generation of subsurface waveguides in diamond.

During preparation of our manuscript, we learned of a concurrent study on direct laser writing of Type II waveguides in diamond which includes characterization of NV centers within the waveguide. It is shown that the presence of the waveguide does not adversely affect the properties of the NV defect making this a promising platform for quantum technologies~\cite{Eaton2016}.\\


\begin{thebibliography}{39}%
\makeatletter
\providecommand \@ifxundefined [1]{%
 \@ifx{#1\undefined}
}%
\providecommand \@ifnum [1]{%
 \ifnum #1\expandafter \@firstoftwo
 \else \expandafter \@secondoftwo
 \fi
}%
\providecommand \@ifx [1]{%
 \ifx #1\expandafter \@firstoftwo
 \else \expandafter \@secondoftwo
 \fi
}%
\providecommand \natexlab [1]{#1}%
\providecommand \enquote  [1]{``#1''}%
\providecommand \bibnamefont  [1]{#1}%
\providecommand \bibfnamefont [1]{#1}%
\providecommand \citenamefont [1]{#1}%
\providecommand \href@noop [0]{\@secondoftwo}%
\providecommand \href [0]{\begingroup \@sanitize@url \@href}%
\providecommand \@href[1]{\@@startlink{#1}\@@href}%
\providecommand \@@href[1]{\endgroup#1\@@endlink}%
\providecommand \@sanitize@url [0]{\catcode `\\12\catcode `\$12\catcode
  `\&12\catcode `\#12\catcode `\^12\catcode `\_12\catcode `\%12\relax}%
\providecommand \@@startlink[1]{}%
\providecommand \@@endlink[0]{}%
\providecommand \url  [0]{\begingroup\@sanitize@url \@url }%
\providecommand \@url [1]{\endgroup\@href {#1}{\urlprefix }}%
\providecommand \urlprefix  [0]{URL }%
\providecommand \Eprint [0]{\href }%
\providecommand \doibase [0]{http://dx.doi.org/}%
\providecommand \selectlanguage [0]{\@gobble}%
\providecommand \bibinfo  [0]{\@secondoftwo}%
\providecommand \bibfield  [0]{\@secondoftwo}%
\providecommand \translation [1]{[#1]}%
\providecommand \BibitemOpen [0]{}%
\providecommand \bibitemStop [0]{}%
\providecommand \bibitemNoStop [0]{.\EOS\space}%
\providecommand \EOS [0]{\spacefactor3000\relax}%
\providecommand \BibitemShut  [1]{\csname bibitem#1\endcsname}%
\let\auto@bib@innerbib\@empty
\bibitem [{\citenamefont {Aharonovich}, \citenamefont {Greentree},\ and\
  \citenamefont {Prawer}(2011)}]{Aharonovich2011}%
  \BibitemOpen
  \bibfield  {author} {\bibinfo {author} {\bibfnamefont {I.}~\bibnamefont
  {Aharonovich}}, \bibinfo {author} {\bibfnamefont {A.~D.}\ \bibnamefont
  {Greentree}}, \ and\ \bibinfo {author} {\bibfnamefont {S.}~\bibnamefont
  {Prawer}},\ }\href {\doibase 10.1038/nphoton.2011.54} {\bibfield  {journal}
  {\bibinfo  {journal} {Nature Photonics}\ }\textbf {\bibinfo {volume} {5}},\
  \bibinfo {pages} {397} (\bibinfo {year} {2011})}\BibitemShut {NoStop}%
\bibitem [{\citenamefont {Jelezko}\ and\ \citenamefont
  {Wrachtrup}(2006)}]{Jelezko2006}%
  \BibitemOpen
  \bibfield  {author} {\bibinfo {author} {\bibfnamefont {F.}~\bibnamefont
  {Jelezko}}\ and\ \bibinfo {author} {\bibfnamefont {J.}~\bibnamefont
  {Wrachtrup}},\ }\href {\doibase 10.1002/pssa.200671403} {\bibfield  {journal}
  {\bibinfo  {journal} {Physica Status Solidi (a)}\ }\textbf {\bibinfo {volume}
  {203}},\ \bibinfo {pages} {3207} (\bibinfo {year} {2006})}\BibitemShut
  {NoStop}%
\bibitem [{\citenamefont {Awschalom}, \citenamefont {Epstein},\ and\
  \citenamefont {Hanson}(2007)}]{Awschalom2007}%
  \BibitemOpen
  \bibfield  {author} {\bibinfo {author} {\bibfnamefont {D.~D.}\ \bibnamefont
  {Awschalom}}, \bibinfo {author} {\bibfnamefont {R.}~\bibnamefont {Epstein}},
  \ and\ \bibinfo {author} {\bibfnamefont {R.}~\bibnamefont {Hanson}},\
  }\href@noop {} {\bibfield  {journal} {\bibinfo  {journal} {Sci. Am.}\
  }\textbf {\bibinfo {volume} {297}},\ \bibinfo {pages} {84} (\bibinfo {year}
  {2007})}\BibitemShut {NoStop}%
\bibitem [{\citenamefont {Bernien}\ \emph {et~al.}(2013)\citenamefont
  {Bernien}, \citenamefont {Hensen}, \citenamefont {Pfaff}, \citenamefont
  {Koolstra}, \citenamefont {Blok}, \citenamefont {Robledo}, \citenamefont
  {Taminiau}, \citenamefont {Markham}, \citenamefont {Twitchen}, \citenamefont
  {Childress},\ and\ \citenamefont {Hanson}}]{Bernien2013}%
  \BibitemOpen
  \bibfield  {author} {\bibinfo {author} {\bibfnamefont {H.}~\bibnamefont
  {Bernien}}, \bibinfo {author} {\bibfnamefont {B.}~\bibnamefont {Hensen}},
  \bibinfo {author} {\bibfnamefont {W.}~\bibnamefont {Pfaff}}, \bibinfo
  {author} {\bibfnamefont {G.}~\bibnamefont {Koolstra}}, \bibinfo {author}
  {\bibfnamefont {M.}~\bibnamefont {Blok}}, \bibinfo {author} {\bibfnamefont
  {L.}~\bibnamefont {Robledo}}, \bibinfo {author} {\bibfnamefont
  {T.}~\bibnamefont {Taminiau}}, \bibinfo {author} {\bibfnamefont
  {M.}~\bibnamefont {Markham}}, \bibinfo {author} {\bibfnamefont
  {D.}~\bibnamefont {Twitchen}}, \bibinfo {author} {\bibfnamefont
  {L.}~\bibnamefont {Childress}}, \ and\ \bibinfo {author} {\bibfnamefont
  {R.}~\bibnamefont {Hanson}},\ }\href@noop {} {\bibfield  {journal} {\bibinfo
  {journal} {Nature}\ }\textbf {\bibinfo {volume} {497}},\ \bibinfo {pages}
  {86} (\bibinfo {year} {2013})}\BibitemShut {NoStop}%
\bibitem [{\citenamefont {Balasubramanian}\ \emph {et~al.}(2008)\citenamefont
  {Balasubramanian}, \citenamefont {Chan}, \citenamefont {Kolesov},
  \citenamefont {Al-Hmoud}, \citenamefont {Tisler}, \citenamefont {Shin},
  \citenamefont {Kim}, \citenamefont {Wojcik}, \citenamefont {Hemmer},
  \citenamefont {Krueger}, \citenamefont {Hanke}, \citenamefont
  {Leitenstorfer}, \citenamefont {Bratschitsch}, \citenamefont {Jelezko},\ and\
  \citenamefont {Wrachtrup}}]{Balsub2008}%
  \BibitemOpen
  \bibfield  {author} {\bibinfo {author} {\bibfnamefont {G.}~\bibnamefont
  {Balasubramanian}}, \bibinfo {author} {\bibfnamefont {I.}~\bibnamefont
  {Chan}}, \bibinfo {author} {\bibfnamefont {R.}~\bibnamefont {Kolesov}},
  \bibinfo {author} {\bibfnamefont {M.}~\bibnamefont {Al-Hmoud}}, \bibinfo
  {author} {\bibfnamefont {J.}~\bibnamefont {Tisler}}, \bibinfo {author}
  {\bibfnamefont {C.}~\bibnamefont {Shin}}, \bibinfo {author} {\bibfnamefont
  {C.}~\bibnamefont {Kim}}, \bibinfo {author} {\bibfnamefont {A.}~\bibnamefont
  {Wojcik}}, \bibinfo {author} {\bibfnamefont {P.~R.}\ \bibnamefont {Hemmer}},
  \bibinfo {author} {\bibfnamefont {A.}~\bibnamefont {Krueger}}, \bibinfo
  {author} {\bibfnamefont {T.}~\bibnamefont {Hanke}}, \bibinfo {author}
  {\bibfnamefont {A.}~\bibnamefont {Leitenstorfer}}, \bibinfo {author}
  {\bibfnamefont {R.}~\bibnamefont {Bratschitsch}}, \bibinfo {author}
  {\bibfnamefont {F.}~\bibnamefont {Jelezko}}, \ and\ \bibinfo {author}
  {\bibfnamefont {J.}~\bibnamefont {Wrachtrup}},\ }\href@noop {} {\bibfield
  {journal} {\bibinfo  {journal} {Nature}\ }\textbf {\bibinfo {volume} {455}},\
  \bibinfo {pages} {648} (\bibinfo {year} {2008})}\BibitemShut {NoStop}%
\bibitem [{\citenamefont {Maze}\ \emph {et~al.}(2008)\citenamefont {Maze},
  \citenamefont {Stanwix}, \citenamefont {Hodges}, \citenamefont {Hong},
  \citenamefont {Taylor}, \citenamefont {Cappellaro}, \citenamefont {Jiang},
  \citenamefont {Dutt}, \citenamefont {Togan}, \citenamefont {Zibrov},
  \citenamefont {Yacoby}, \citenamefont {Walsworth},\ and\ \citenamefont
  {Lukin}}]{Maze2008}%
  \BibitemOpen
  \bibfield  {author} {\bibinfo {author} {\bibfnamefont {J.~R.}\ \bibnamefont
  {Maze}}, \bibinfo {author} {\bibfnamefont {P.~L.}\ \bibnamefont {Stanwix}},
  \bibinfo {author} {\bibfnamefont {J.~S.}\ \bibnamefont {Hodges}}, \bibinfo
  {author} {\bibfnamefont {S.}~\bibnamefont {Hong}}, \bibinfo {author}
  {\bibfnamefont {J.~M.}\ \bibnamefont {Taylor}}, \bibinfo {author}
  {\bibfnamefont {P.}~\bibnamefont {Cappellaro}}, \bibinfo {author}
  {\bibfnamefont {L.}~\bibnamefont {Jiang}}, \bibinfo {author} {\bibfnamefont
  {M.~V.~G.}\ \bibnamefont {Dutt}}, \bibinfo {author} {\bibfnamefont
  {E.}~\bibnamefont {Togan}}, \bibinfo {author} {\bibfnamefont {A.~S.}\
  \bibnamefont {Zibrov}}, \bibinfo {author} {\bibfnamefont {A.}~\bibnamefont
  {Yacoby}}, \bibinfo {author} {\bibfnamefont {R.~L.}\ \bibnamefont
  {Walsworth}}, \ and\ \bibinfo {author} {\bibfnamefont {M.~D.}\ \bibnamefont
  {Lukin}},\ }\href@noop {} {\bibfield  {journal} {\bibinfo  {journal}
  {Nature}\ }\textbf {\bibinfo {volume} {455}},\ \bibinfo {pages} {644}
  (\bibinfo {year} {2008})}\BibitemShut {NoStop}%
\bibitem [{\citenamefont {Neumann}\ \emph {et~al.}(2013)\citenamefont
  {Neumann}, \citenamefont {Jakobi}, \citenamefont {Dolde}, \citenamefont
  {Burk}, \citenamefont {Reuter}, \citenamefont {Waldherr}, \citenamefont
  {Honert}, \citenamefont {Wolf}, \citenamefont {Brunner}, \citenamefont
  {Shim}, \citenamefont {Sumiya}, \citenamefont {Isoya},\ and\ \citenamefont
  {Wrachtrup}}]{Neumann2013}%
  \BibitemOpen
  \bibfield  {author} {\bibinfo {author} {\bibfnamefont {P.}~\bibnamefont
  {Neumann}}, \bibinfo {author} {\bibfnamefont {I.}~\bibnamefont {Jakobi}},
  \bibinfo {author} {\bibfnamefont {F.}~\bibnamefont {Dolde}}, \bibinfo
  {author} {\bibfnamefont {C.}~\bibnamefont {Burk}}, \bibinfo {author}
  {\bibfnamefont {R.}~\bibnamefont {Reuter}}, \bibinfo {author} {\bibfnamefont
  {G.}~\bibnamefont {Waldherr}}, \bibinfo {author} {\bibfnamefont
  {J.}~\bibnamefont {Honert}}, \bibinfo {author} {\bibfnamefont
  {T.}~\bibnamefont {Wolf}}, \bibinfo {author} {\bibfnamefont {A.}~\bibnamefont
  {Brunner}}, \bibinfo {author} {\bibfnamefont {J.~H.}\ \bibnamefont {Shim}},
  \bibinfo {author} {\bibfnamefont {D.~S.~H.}\ \bibnamefont {Sumiya}}, \bibinfo
  {author} {\bibfnamefont {J.}~\bibnamefont {Isoya}}, \ and\ \bibinfo {author}
  {\bibfnamefont {J.}~\bibnamefont {Wrachtrup}},\ }\href@noop {} {\bibfield
  {journal} {\bibinfo  {journal} {Nano Lett}\ }\textbf {\bibinfo {volume}
  {13}},\ \bibinfo {pages} {2378} (\bibinfo {year} {2013})}\BibitemShut
  {NoStop}%
\bibitem [{\citenamefont {Mildren}\ and\ \citenamefont
  {Sabella}(2009)}]{Mildren2009}%
  \BibitemOpen
  \bibfield  {author} {\bibinfo {author} {\bibfnamefont {R.~P.}\ \bibnamefont
  {Mildren}}\ and\ \bibinfo {author} {\bibfnamefont {A.}~\bibnamefont
  {Sabella}},\ }\href@noop {} {\bibfield  {journal} {\bibinfo  {journal}
  {Optics Letters}\ }\textbf {\bibinfo {volume} {34}},\ \bibinfo {pages} {2811}
  (\bibinfo {year} {2009})}\BibitemShut {NoStop}%
\bibitem [{\citenamefont {Reilly}\ \emph {et~al.}(2015)\citenamefont {Reilly},
  \citenamefont {Savitski}, \citenamefont {Liu}, \citenamefont {Gu},
  \citenamefont {Dawson},\ and\ \citenamefont {Kemp}}]{Kemp2015}%
  \BibitemOpen
  \bibfield  {author} {\bibinfo {author} {\bibfnamefont {S.}~\bibnamefont
  {Reilly}}, \bibinfo {author} {\bibfnamefont {V.~G.}\ \bibnamefont
  {Savitski}}, \bibinfo {author} {\bibfnamefont {H.}~\bibnamefont {Liu}},
  \bibinfo {author} {\bibfnamefont {E.}~\bibnamefont {Gu}}, \bibinfo {author}
  {\bibfnamefont {M.~D.}\ \bibnamefont {Dawson}}, \ and\ \bibinfo {author}
  {\bibfnamefont {A.~J.}\ \bibnamefont {Kemp}},\ }\href@noop {} {\bibfield
  {journal} {\bibinfo  {journal} {Opt. Lett.}\ }\textbf {\bibinfo {volume}
  {40}},\ \bibinfo {pages} {930} (\bibinfo {year} {2015})}\BibitemShut
  {NoStop}%
\bibitem [{\citenamefont {Poem}\ \emph {et~al.}(2016)\citenamefont {Poem},
  \citenamefont {Weinzetl}, \citenamefont {Klatzow}, \citenamefont {Kaczmarek},
  \citenamefont {Munns}, \citenamefont {Champion}, \citenamefont {Saunders},
  \citenamefont {Nunn},\ and\ \citenamefont {Walmsley}}]{eilon2016}%
  \BibitemOpen
  \bibfield  {author} {\bibinfo {author} {\bibfnamefont {E.}~\bibnamefont
  {Poem}}, \bibinfo {author} {\bibfnamefont {C.}~\bibnamefont {Weinzetl}},
  \bibinfo {author} {\bibfnamefont {J.}~\bibnamefont {Klatzow}}, \bibinfo
  {author} {\bibfnamefont {K.}~\bibnamefont {Kaczmarek}}, \bibinfo {author}
  {\bibfnamefont {J.}~\bibnamefont {Munns}}, \bibinfo {author} {\bibfnamefont
  {T.}~\bibnamefont {Champion}}, \bibinfo {author} {\bibfnamefont
  {D.}~\bibnamefont {Saunders}}, \bibinfo {author} {\bibfnamefont
  {J.}~\bibnamefont {Nunn}}, \ and\ \bibinfo {author} {\bibfnamefont
  {I.}~\bibnamefont {Walmsley}},\ }\href@noop {} {\bibfield  {journal}
  {\bibinfo  {journal} {Phys. Rev. B}\ }\textbf {\bibinfo {volume} {91}},\
  \bibinfo {pages} {205108} (\bibinfo {year} {2016})}\BibitemShut {NoStop}%
\bibitem [{\citenamefont {Picollo}\ \emph {et~al.}(2013)\citenamefont
  {Picollo}, \citenamefont {Gosso}, \citenamefont {Vittone}, \citenamefont
  {Pasquarelli}, \citenamefont {Carbone}, \citenamefont {Olivero},\ and\
  \citenamefont {Carabelli}}]{Picollo2013}%
  \BibitemOpen
  \bibfield  {author} {\bibinfo {author} {\bibfnamefont {F.}~\bibnamefont
  {Picollo}}, \bibinfo {author} {\bibfnamefont {S.}~\bibnamefont {Gosso}},
  \bibinfo {author} {\bibfnamefont {E.}~\bibnamefont {Vittone}}, \bibinfo
  {author} {\bibfnamefont {A.}~\bibnamefont {Pasquarelli}}, \bibinfo {author}
  {\bibfnamefont {E.}~\bibnamefont {Carbone}}, \bibinfo {author} {\bibfnamefont
  {P.}~\bibnamefont {Olivero}}, \ and\ \bibinfo {author} {\bibfnamefont
  {V.}~\bibnamefont {Carabelli}},\ }\href {\doibase 10.1002/adma.201300710}
  {\bibfield  {journal} {\bibinfo  {journal} {Advanced materials (Deerfield
  Beach, Fla.)}\ }\textbf {\bibinfo {volume} {25}},\ \bibinfo {pages} {4696}
  (\bibinfo {year} {2013})}\BibitemShut {NoStop}%
\bibitem [{\citenamefont {Clevenson}\ \emph {et~al.}(2015)\citenamefont
  {Clevenson}, \citenamefont {Trusheim}, \citenamefont {Teale}, \citenamefont
  {Schröder}, \citenamefont {Braje},\ and\ \citenamefont
  {Englund}}]{clevenson2015}%
  \BibitemOpen
  \bibfield  {author} {\bibinfo {author} {\bibfnamefont {H.}~\bibnamefont
  {Clevenson}}, \bibinfo {author} {\bibfnamefont {M.~E.}\ \bibnamefont
  {Trusheim}}, \bibinfo {author} {\bibfnamefont {C.}~\bibnamefont {Teale}},
  \bibinfo {author} {\bibfnamefont {T.}~\bibnamefont {Schröder}}, \bibinfo
  {author} {\bibfnamefont {D.}~\bibnamefont {Braje}}, \ and\ \bibinfo {author}
  {\bibfnamefont {D.}~\bibnamefont {Englund}},\ }\href@noop {} {\bibfield
  {journal} {\bibinfo  {journal} {Nat. Physics}\ }\textbf {\bibinfo {volume}
  {11}},\ \bibinfo {pages} {393} (\bibinfo {year} {2015})}\BibitemShut
  {NoStop}%
\bibitem [{\citenamefont {Hiscocks}\ \emph {et~al.}(2008)\citenamefont
  {Hiscocks}, \citenamefont {Ganesan}, \citenamefont {Gibson}, \citenamefont
  {Huntington}, \citenamefont {Ladouceur},\ and\ \citenamefont
  {Prawer}}]{Prawer2008}%
  \BibitemOpen
  \bibfield  {author} {\bibinfo {author} {\bibfnamefont {M.~P.}\ \bibnamefont
  {Hiscocks}}, \bibinfo {author} {\bibfnamefont {K.}~\bibnamefont {Ganesan}},
  \bibinfo {author} {\bibfnamefont {B.~C.}\ \bibnamefont {Gibson}}, \bibinfo
  {author} {\bibfnamefont {S.~T.}\ \bibnamefont {Huntington}}, \bibinfo
  {author} {\bibfnamefont {F.}~\bibnamefont {Ladouceur}}, \ and\ \bibinfo
  {author} {\bibfnamefont {S.}~\bibnamefont {Prawer}},\ }\href@noop {}
  {\bibfield  {journal} {\bibinfo  {journal} {Optics Express}\ }\textbf
  {\bibinfo {volume} {16}},\ \bibinfo {pages} {19512} (\bibinfo {year}
  {2008})}\BibitemShut {NoStop}%
\bibitem [{\citenamefont {Zhang}\ \emph {et~al.}(2011)\citenamefont {Zhang},
  \citenamefont {McKnight}, \citenamefont {Tian}, \citenamefont {Calvez},
  \citenamefont {Gu},\ and\ \citenamefont {Dawson}}]{Erdan2011}%
  \BibitemOpen
  \bibfield  {author} {\bibinfo {author} {\bibfnamefont {Y.}~\bibnamefont
  {Zhang}}, \bibinfo {author} {\bibfnamefont {L.}~\bibnamefont {McKnight}},
  \bibinfo {author} {\bibfnamefont {Z.}~\bibnamefont {Tian}}, \bibinfo {author}
  {\bibfnamefont {S.}~\bibnamefont {Calvez}}, \bibinfo {author} {\bibfnamefont
  {E.}~\bibnamefont {Gu}}, \ and\ \bibinfo {author} {\bibfnamefont {M.~D.}\
  \bibnamefont {Dawson}},\ }\href@noop {} {\bibfield  {journal} {\bibinfo
  {journal} {Diam. Rel. Mat.}\ }\textbf {\bibinfo {volume} {20}},\ \bibinfo
  {pages} {564} (\bibinfo {year} {2011})}\BibitemShut {NoStop}%
\bibitem [{\citenamefont {Burek}\ \emph {et~al.}(2014)\citenamefont {Burek},
  \citenamefont {Chu}, \citenamefont {Liddy}, \citenamefont {Patel},
  \citenamefont {Rochman}, \citenamefont {Meesala}, \citenamefont {Hong},
  \citenamefont {Quan}, \citenamefont {Lukin},\ and\ \citenamefont
  {Loncar}}]{Burek2014}%
  \BibitemOpen
  \bibfield  {author} {\bibinfo {author} {\bibfnamefont {M.~J.}\ \bibnamefont
  {Burek}}, \bibinfo {author} {\bibfnamefont {Y.}~\bibnamefont {Chu}}, \bibinfo
  {author} {\bibfnamefont {M.~S.}\ \bibnamefont {Liddy}}, \bibinfo {author}
  {\bibfnamefont {P.}~\bibnamefont {Patel}}, \bibinfo {author} {\bibfnamefont
  {J.}~\bibnamefont {Rochman}}, \bibinfo {author} {\bibfnamefont
  {S.}~\bibnamefont {Meesala}}, \bibinfo {author} {\bibfnamefont
  {W.}~\bibnamefont {Hong}}, \bibinfo {author} {\bibfnamefont {Q.}~\bibnamefont
  {Quan}}, \bibinfo {author} {\bibfnamefont {M.~D.}\ \bibnamefont {Lukin}}, \
  and\ \bibinfo {author} {\bibfnamefont {M.}~\bibnamefont {Loncar}},\
  }\href@noop {} {\bibfield  {journal} {\bibinfo  {journal} {Nat. Comms.}\
  }\textbf {\bibinfo {volume} {5}},\ \bibinfo {pages} {5718} (\bibinfo {year}
  {2014})}\BibitemShut {NoStop}%
\bibitem [{\citenamefont {Khanaliloo}\ \emph {et~al.}(2015)\citenamefont
  {Khanaliloo}, \citenamefont {Jayakumar}, \citenamefont {Hryciw},
  \citenamefont {Lake}, \citenamefont {Kaviani},\ and\ \citenamefont
  {Barclay}}]{barclay2015}%
  \BibitemOpen
  \bibfield  {author} {\bibinfo {author} {\bibfnamefont {B.}~\bibnamefont
  {Khanaliloo}}, \bibinfo {author} {\bibfnamefont {H.}~\bibnamefont
  {Jayakumar}}, \bibinfo {author} {\bibfnamefont {A.~C.}\ \bibnamefont
  {Hryciw}}, \bibinfo {author} {\bibfnamefont {D.~P.}\ \bibnamefont {Lake}},
  \bibinfo {author} {\bibfnamefont {H.}~\bibnamefont {Kaviani}}, \ and\
  \bibinfo {author} {\bibfnamefont {P.~E.}\ \bibnamefont {Barclay}},\
  }\href@noop {} {\bibfield  {journal} {\bibinfo  {journal} {Phys. Rev. X}\
  }\textbf {\bibinfo {volume} {5}},\ \bibinfo {pages} {041051} (\bibinfo {year}
  {2015})}\BibitemShut {NoStop}%
\bibitem [{\citenamefont {Lagomarsino}\ \emph {et~al.}(2010)\citenamefont
  {Lagomarsino}, \citenamefont {Olivero}, \citenamefont {Bosia}, \citenamefont
  {Vannoni}, \citenamefont {Calusi}, \citenamefont {Giuntini},\ and\
  \citenamefont {Massi}}]{Lagomarsino2010}%
  \BibitemOpen
  \bibfield  {author} {\bibinfo {author} {\bibfnamefont {S.}~\bibnamefont
  {Lagomarsino}}, \bibinfo {author} {\bibfnamefont {P.}~\bibnamefont
  {Olivero}}, \bibinfo {author} {\bibfnamefont {F.}~\bibnamefont {Bosia}},
  \bibinfo {author} {\bibfnamefont {M.}~\bibnamefont {Vannoni}}, \bibinfo
  {author} {\bibfnamefont {S.}~\bibnamefont {Calusi}}, \bibinfo {author}
  {\bibfnamefont {L.}~\bibnamefont {Giuntini}}, \ and\ \bibinfo {author}
  {\bibfnamefont {M.}~\bibnamefont {Massi}},\ }\href@noop {} {\bibfield
  {journal} {\bibinfo  {journal} {{Phys. Rev. Lett.}}\ }\textbf {\bibinfo
  {volume} {{105}}},\ \bibinfo {pages} {233903} (\bibinfo {year}
  {2010})}\BibitemShut {NoStop}%
\bibitem [{\citenamefont {Chen}\ and\ \citenamefont
  {de~Aldana}(2014)}]{Chen2014}%
  \BibitemOpen
  \bibfield  {author} {\bibinfo {author} {\bibfnamefont {F.}~\bibnamefont
  {Chen}}\ and\ \bibinfo {author} {\bibfnamefont {J.~V.}\ \bibnamefont
  {de~Aldana}},\ }\href@noop {} {\bibfield  {journal} {\bibinfo  {journal}
  {Laser Photonics Rev.}\ }\textbf {\bibinfo {volume} {8}},\ \bibinfo {pages}
  {251} (\bibinfo {year} {2014})}\BibitemShut {NoStop}%
\bibitem [{\citenamefont {Gross}, \citenamefont {Dubov},\ and\ \citenamefont
  {Withford}(2015)}]{Gross2015}%
  \BibitemOpen
  \bibfield  {author} {\bibinfo {author} {\bibfnamefont {S.}~\bibnamefont
  {Gross}}, \bibinfo {author} {\bibfnamefont {M.}~\bibnamefont {Dubov}}, \ and\
  \bibinfo {author} {\bibfnamefont {M.}~\bibnamefont {Withford}},\ }\href@noop
  {} {\bibfield  {journal} {\bibinfo  {journal} {Opt. Express}\ }\textbf
  {\bibinfo {volume} {23}},\ \bibinfo {pages} {7767} (\bibinfo {year}
  {2015})}\BibitemShut {NoStop}%
\bibitem [{\citenamefont {Osellame}\ \emph {et~al.}(2007)\citenamefont
  {Osellame}, \citenamefont {Lobino}, \citenamefont {Chiodo}, \citenamefont
  {Marangoni}, \citenamefont {Cerullo}, \citenamefont {Ramponi}, \citenamefont
  {Bookey}, \citenamefont {Thomson}, \citenamefont {Psaila},\ and\
  \citenamefont {Kar}}]{osellame07b}%
  \BibitemOpen
  \bibfield  {author} {\bibinfo {author} {\bibfnamefont {R.}~\bibnamefont
  {Osellame}}, \bibinfo {author} {\bibfnamefont {M.}~\bibnamefont {Lobino}},
  \bibinfo {author} {\bibfnamefont {N.}~\bibnamefont {Chiodo}}, \bibinfo
  {author} {\bibfnamefont {M.}~\bibnamefont {Marangoni}}, \bibinfo {author}
  {\bibfnamefont {G.}~\bibnamefont {Cerullo}}, \bibinfo {author} {\bibfnamefont
  {R.}~\bibnamefont {Ramponi}}, \bibinfo {author} {\bibfnamefont
  {H.}~\bibnamefont {Bookey}}, \bibinfo {author} {\bibfnamefont
  {R.}~\bibnamefont {Thomson}}, \bibinfo {author} {\bibfnamefont
  {N.}~\bibnamefont {Psaila}}, \ and\ \bibinfo {author} {\bibfnamefont
  {A.}~\bibnamefont {Kar}},\ }\href@noop {} {\bibfield  {journal} {\bibinfo
  {journal} {Appl. Phys. Lett.}\ }\textbf {\bibinfo {volume} {90}},\ \bibinfo
  {pages} {241107} (\bibinfo {year} {2007})}\BibitemShut {NoStop}%
\bibitem [{\citenamefont {Thomas}\ \emph {et~al.}(2011)\citenamefont {Thomas},
  \citenamefont {Heinrich}, \citenamefont {Zeil}, \citenamefont {Hilbert},
  \citenamefont {Rademaker}, \citenamefont {Riedel}, \citenamefont {Ringleb},
  \citenamefont {Dubs}, \citenamefont {Ruske}, \citenamefont {Nolte},\ and\
  \citenamefont {T\"{u}nnermann}}]{thomas2011}%
  \BibitemOpen
  \bibfield  {author} {\bibinfo {author} {\bibfnamefont {J.}~\bibnamefont
  {Thomas}}, \bibinfo {author} {\bibfnamefont {M.}~\bibnamefont {Heinrich}},
  \bibinfo {author} {\bibfnamefont {P.}~\bibnamefont {Zeil}}, \bibinfo {author}
  {\bibfnamefont {V.}~\bibnamefont {Hilbert}}, \bibinfo {author} {\bibfnamefont
  {K.}~\bibnamefont {Rademaker}}, \bibinfo {author} {\bibfnamefont
  {R.}~\bibnamefont {Riedel}}, \bibinfo {author} {\bibfnamefont
  {S.}~\bibnamefont {Ringleb}}, \bibinfo {author} {\bibfnamefont
  {C.}~\bibnamefont {Dubs}}, \bibinfo {author} {\bibfnamefont {J.-P.}\
  \bibnamefont {Ruske}}, \bibinfo {author} {\bibfnamefont {S.}~\bibnamefont
  {Nolte}}, \ and\ \bibinfo {author} {\bibfnamefont {A.}~\bibnamefont
  {T\"{u}nnermann}},\ }\href {\doibase 10.1002/pssa.201026452} {\bibfield
  {journal} {\bibinfo  {journal} {Physica Status Solidi (a)}\ }\textbf
  {\bibinfo {volume} {208}},\ \bibinfo {pages} {276} (\bibinfo {year}
  {2011})}\BibitemShut {NoStop}%
\bibitem [{\citenamefont {Huang}\ \emph {et~al.}(2015)\citenamefont {Huang},
  \citenamefont {Salter}, \citenamefont {Karpinski}, \citenamefont {Smith},
  \citenamefont {Payne},\ and\ \citenamefont {Booth}}]{huang2015}%
  \BibitemOpen
  \bibfield  {author} {\bibinfo {author} {\bibfnamefont {L.}~\bibnamefont
  {Huang}}, \bibinfo {author} {\bibfnamefont {P.}~\bibnamefont {Salter}},
  \bibinfo {author} {\bibfnamefont {M.}~\bibnamefont {Karpinski}}, \bibinfo
  {author} {\bibfnamefont {B.}~\bibnamefont {Smith}}, \bibinfo {author}
  {\bibfnamefont {F.}~\bibnamefont {Payne}}, \ and\ \bibinfo {author}
  {\bibfnamefont {M.}~\bibnamefont {Booth}},\ }\href@noop {} {\bibfield
  {journal} {\bibinfo  {journal} {Appl. Phys. A}\ }\textbf {\bibinfo {volume}
  {118}},\ \bibinfo {pages} {831} (\bibinfo {year} {2015})}\BibitemShut
  {NoStop}%
\bibitem [{\citenamefont {Campbell}\ \emph {et~al.}(2007)\citenamefont
  {Campbell}, \citenamefont {Thomson}, \citenamefont {Hand}, \citenamefont
  {Kar}, \citenamefont {Reid}, \citenamefont {Canalias}, \citenamefont
  {Pasiskevicius},\ and\ \citenamefont {Laurell}}]{campbell2007b}%
  \BibitemOpen
  \bibfield  {author} {\bibinfo {author} {\bibfnamefont {S.}~\bibnamefont
  {Campbell}}, \bibinfo {author} {\bibfnamefont {R.~R.}\ \bibnamefont
  {Thomson}}, \bibinfo {author} {\bibfnamefont {D.~P.}\ \bibnamefont {Hand}},
  \bibinfo {author} {\bibfnamefont {A.~K.}\ \bibnamefont {Kar}}, \bibinfo
  {author} {\bibfnamefont {D.~T.}\ \bibnamefont {Reid}}, \bibinfo {author}
  {\bibfnamefont {C.}~\bibnamefont {Canalias}}, \bibinfo {author}
  {\bibfnamefont {V.}~\bibnamefont {Pasiskevicius}}, \ and\ \bibinfo {author}
  {\bibfnamefont {F.}~\bibnamefont {Laurell}},\ }\href@noop {} {\bibfield
  {journal} {\bibinfo  {journal} {Optics Express}\ }\textbf {\bibinfo {volume}
  {15}},\ \bibinfo {pages} {17146} (\bibinfo {year} {2007})}\BibitemShut
  {NoStop}%
\bibitem [{\citenamefont {Bai}\ \emph {et~al.}(2012)\citenamefont {Bai},
  \citenamefont {Cheng}, \citenamefont {Long}, \citenamefont {Wang},
  \citenamefont {Zhao}, \citenamefont {Chen}, \citenamefont {Stoian},\ and\
  \citenamefont {Hui}}]{stoian2012}%
  \BibitemOpen
  \bibfield  {author} {\bibinfo {author} {\bibfnamefont {J.}~\bibnamefont
  {Bai}}, \bibinfo {author} {\bibfnamefont {G.}~\bibnamefont {Cheng}}, \bibinfo
  {author} {\bibfnamefont {X.}~\bibnamefont {Long}}, \bibinfo {author}
  {\bibfnamefont {Y.}~\bibnamefont {Wang}}, \bibinfo {author} {\bibfnamefont
  {W.}~\bibnamefont {Zhao}}, \bibinfo {author} {\bibfnamefont {G.}~\bibnamefont
  {Chen}}, \bibinfo {author} {\bibfnamefont {R.}~\bibnamefont {Stoian}}, \ and\
  \bibinfo {author} {\bibfnamefont {R.}~\bibnamefont {Hui}},\ }\href@noop {}
  {\bibfield  {journal} {\bibinfo  {journal} {Opt. Express}\ }\textbf {\bibinfo
  {volume} {20}},\ \bibinfo {pages} {15035} (\bibinfo {year}
  {2012})}\BibitemShut {NoStop}%
\bibitem [{\citenamefont {Lagomarsino}\ \emph {et~al.}(2016)\citenamefont
  {Lagomarsino}, \citenamefont {Sciortino}, \citenamefont {Obreshkov},
  \citenamefont {Apostolova}, \citenamefont {Corsi}, \citenamefont {Bellini},
  \citenamefont {Berdermann},\ and\ \citenamefont {Schmidt}}]{Lagomarsino2016}%
  \BibitemOpen
  \bibfield  {author} {\bibinfo {author} {\bibfnamefont {S.}~\bibnamefont
  {Lagomarsino}}, \bibinfo {author} {\bibfnamefont {S.}~\bibnamefont
  {Sciortino}}, \bibinfo {author} {\bibfnamefont {B.}~\bibnamefont
  {Obreshkov}}, \bibinfo {author} {\bibfnamefont {T.}~\bibnamefont
  {Apostolova}}, \bibinfo {author} {\bibfnamefont {C.}~\bibnamefont {Corsi}},
  \bibinfo {author} {\bibfnamefont {M.}~\bibnamefont {Bellini}}, \bibinfo
  {author} {\bibfnamefont {E.}~\bibnamefont {Berdermann}}, \ and\ \bibinfo
  {author} {\bibfnamefont {C.~J.}\ \bibnamefont {Schmidt}},\ }\href@noop {}
  {\bibfield  {journal} {\bibinfo  {journal} {Phys. Rev. B}\ }\textbf {\bibinfo
  {volume} {93}},\ \bibinfo {pages} {085128} (\bibinfo {year}
  {2016})}\BibitemShut {NoStop}%
\bibitem [{\citenamefont {Kononenko}\ \emph {et~al.}(2008)\citenamefont
  {Kononenko}, \citenamefont {Meier}, \citenamefont {Komlenok}, \citenamefont
  {Pimenov}, \citenamefont {Romano}, \citenamefont {Pashinin},\ and\
  \citenamefont {Konov}}]{Kononenko2008}%
  \BibitemOpen
  \bibfield  {author} {\bibinfo {author} {\bibfnamefont {T.~V.}\ \bibnamefont
  {Kononenko}}, \bibinfo {author} {\bibfnamefont {M.}~\bibnamefont {Meier}},
  \bibinfo {author} {\bibfnamefont {M.~S.}\ \bibnamefont {Komlenok}}, \bibinfo
  {author} {\bibfnamefont {S.~M.}\ \bibnamefont {Pimenov}}, \bibinfo {author}
  {\bibfnamefont {V.}~\bibnamefont {Romano}}, \bibinfo {author} {\bibfnamefont
  {V.~P.}\ \bibnamefont {Pashinin}}, \ and\ \bibinfo {author} {\bibfnamefont
  {V.~I.}\ \bibnamefont {Konov}},\ }\href@noop {} {\bibfield  {journal}
  {\bibinfo  {journal} {Applied Physics A}\ }\textbf {\bibinfo {volume} {90}},\
  \bibinfo {pages} {645} (\bibinfo {year} {2008})}\BibitemShut {NoStop}%
\bibitem [{\citenamefont {Salter}\ and\ \citenamefont
  {Booth}(2014)}]{SalterSPIE2014}%
  \BibitemOpen
  \bibfield  {author} {\bibinfo {author} {\bibfnamefont {P.}~\bibnamefont
  {Salter}}\ and\ \bibinfo {author} {\bibfnamefont {M.~J.}\ \bibnamefont
  {Booth}},\ }\href@noop {} {\bibfield  {journal} {\bibinfo  {journal} {Proc.
  SPIE MOEMS-MEMS}\ }\textbf {\bibinfo {volume} {8974}},\ \bibinfo {pages}
  {89740T} (\bibinfo {year} {2014})}\BibitemShut {NoStop}%
\bibitem [{\citenamefont {Pimenov}\ \emph {et~al.}(2016)\citenamefont
  {Pimenov}, \citenamefont {Khomich}, \citenamefont {Neuenshwander},
  \citenamefont {Jaggi},\ and\ \citenamefont {Romano}}]{neuenshwander2016}%
  \BibitemOpen
  \bibfield  {author} {\bibinfo {author} {\bibfnamefont {S.~M.}\ \bibnamefont
  {Pimenov}}, \bibinfo {author} {\bibfnamefont {A.~A.}\ \bibnamefont
  {Khomich}}, \bibinfo {author} {\bibfnamefont {B.}~\bibnamefont
  {Neuenshwander}}, \bibinfo {author} {\bibfnamefont {B.}~\bibnamefont
  {Jaggi}}, \ and\ \bibinfo {author} {\bibfnamefont {V.}~\bibnamefont
  {Romano}},\ }\href@noop {} {\bibfield  {journal} {\bibinfo  {journal} {JOSA
  b}\ }\textbf {\bibinfo {volume} {33}},\ \bibinfo {pages} {B49} (\bibinfo
  {year} {2016})}\BibitemShut {NoStop}%
\bibitem [{\citenamefont {Lagomarsino}\ \emph {et~al.}(2013)\citenamefont
  {Lagomarsino}, \citenamefont {Bellini}, \citenamefont {Corsi}, \citenamefont
  {Gorelli}, \citenamefont {Parrini}, \citenamefont {Santoro},\ and\
  \citenamefont {Sciortino}}]{Lagomarsino2013a}%
  \BibitemOpen
  \bibfield  {author} {\bibinfo {author} {\bibfnamefont {S.}~\bibnamefont
  {Lagomarsino}}, \bibinfo {author} {\bibfnamefont {M.}~\bibnamefont
  {Bellini}}, \bibinfo {author} {\bibfnamefont {C.}~\bibnamefont {Corsi}},
  \bibinfo {author} {\bibfnamefont {F.}~\bibnamefont {Gorelli}}, \bibinfo
  {author} {\bibfnamefont {G.}~\bibnamefont {Parrini}}, \bibinfo {author}
  {\bibfnamefont {M.}~\bibnamefont {Santoro}}, \ and\ \bibinfo {author}
  {\bibfnamefont {S.}~\bibnamefont {Sciortino}},\ }\href {\doibase
  10.1063/1.4839555} {\bibfield  {journal} {\bibinfo  {journal} {Applied
  Physics Letters}\ }\textbf {\bibinfo {volume} {103}},\ \bibinfo {pages}
  {233507} (\bibinfo {year} {2013})}\BibitemShut {NoStop}%
\bibitem{Unpub_FIB}\BibitemOpen {\bibfield  {journal} {\bibinfo  {journal}
  {Manuscript in preparation}\ }}\BibitemShut {NoStop}%
\bibitem [{\citenamefont {Sun}, \citenamefont {Salter},\ and\ \citenamefont
  {Booth}(2014)}]{Sun2014b}%
  \BibitemOpen
  \bibfield  {author} {\bibinfo {author} {\bibfnamefont {B.}~\bibnamefont
  {Sun}}, \bibinfo {author} {\bibfnamefont {P.~S.}\ \bibnamefont {Salter}}, \
  and\ \bibinfo {author} {\bibfnamefont {M.~J.}\ \bibnamefont {Booth}},\ }\href
  {\doibase 10.1063/1.4902998} {\bibfield  {journal} {\bibinfo  {journal}
  {Appl. Phys. Lett}\ }\textbf {\bibinfo {volume} {105}},\ \bibinfo {pages}
  {231105} (\bibinfo {year} {2014})}\BibitemShut {NoStop}%
\bibitem [{\citenamefont {Oh}\ \emph {et~al.}(2013)\citenamefont {Oh},
  \citenamefont {Caylar}, \citenamefont {Pomorski},\ and\ \citenamefont
  {Wengler}}]{Oh2013b}%
  \BibitemOpen
  \bibfield  {author} {\bibinfo {author} {\bibfnamefont {A.}~\bibnamefont
  {Oh}}, \bibinfo {author} {\bibfnamefont {B.}~\bibnamefont {Caylar}}, \bibinfo
  {author} {\bibfnamefont {M.}~\bibnamefont {Pomorski}}, \ and\ \bibinfo
  {author} {\bibfnamefont {T.}~\bibnamefont {Wengler}},\ }\href {\doibase
  10.1016/j.diamond.2013.06.003} {\bibfield  {journal} {\bibinfo  {journal}
  {Diamond and Related Materials}\ }\textbf {\bibinfo {volume} {38}},\ \bibinfo
  {pages} {9} (\bibinfo {year} {2013})}\BibitemShut {NoStop}%
\bibitem [{\citenamefont {Caylar}, \citenamefont {Pomorski},\ and\
  \citenamefont {Bergonzo}(2013)}]{Caylar2013a}%
  \BibitemOpen
  \bibfield  {author} {\bibinfo {author} {\bibfnamefont {B.}~\bibnamefont
  {Caylar}}, \bibinfo {author} {\bibfnamefont {M.}~\bibnamefont {Pomorski}}, \
  and\ \bibinfo {author} {\bibfnamefont {P.}~\bibnamefont {Bergonzo}},\ }\href
  {\doibase 10.1063/1.4816328} {\bibfield  {journal} {\bibinfo  {journal}
  {Applied Physics Letters}\ }\textbf {\bibinfo {volume} {103}},\ \bibinfo
  {pages} {043504} (\bibinfo {year} {2013})}\BibitemShut {NoStop}%
\bibitem [{\citenamefont {Kononenko}\ \emph {et~al.}(2013)\citenamefont
  {Kononenko}, \citenamefont {Ralchenko}, \citenamefont {Bolshakov},
  \citenamefont {Konov}, \citenamefont {Allegrini}, \citenamefont {Pacilli},
  \citenamefont {Conte},\ and\ \citenamefont {Spiriti}}]{Kononenko2013e}%
  \BibitemOpen
  \bibfield  {author} {\bibinfo {author} {\bibfnamefont {T.}~\bibnamefont
  {Kononenko}}, \bibinfo {author} {\bibfnamefont {V.}~\bibnamefont
  {Ralchenko}}, \bibinfo {author} {\bibfnamefont {A.}~\bibnamefont
  {Bolshakov}}, \bibinfo {author} {\bibfnamefont {V.}~\bibnamefont {Konov}},
  \bibinfo {author} {\bibfnamefont {P.}~\bibnamefont {Allegrini}}, \bibinfo
  {author} {\bibfnamefont {M.}~\bibnamefont {Pacilli}}, \bibinfo {author}
  {\bibfnamefont {G.}~\bibnamefont {Conte}}, \ and\ \bibinfo {author}
  {\bibfnamefont {E.}~\bibnamefont {Spiriti}},\ }\href {\doibase
  10.1007/s00339-013-8091-7} {\bibfield  {journal} {\bibinfo  {journal}
  {Applied Physics A}\ }\textbf {\bibinfo {volume} {114}},\ \bibinfo {pages}
  {297} (\bibinfo {year} {2013})}\BibitemShut {NoStop}%
\bibitem [{\citenamefont {Huang}\ \emph {et~al.}(2016)\citenamefont {Huang},
  \citenamefont {Salter}, \citenamefont {Payne},\ and\ \citenamefont
  {Booth}}]{huang2016}%
  \BibitemOpen
  \bibfield  {author} {\bibinfo {author} {\bibfnamefont {L.}~\bibnamefont
  {Huang}}, \bibinfo {author} {\bibfnamefont {P.}~\bibnamefont {Salter}},
  \bibinfo {author} {\bibfnamefont {F.}~\bibnamefont {Payne}}, \ and\ \bibinfo
  {author} {\bibfnamefont {M.}~\bibnamefont {Booth}},\ }\href@noop {}
  {\bibfield  {journal} {\bibinfo  {journal} {Opt. Express}\ }\textbf {\bibinfo
  {volume} {24}},\ \bibinfo {pages} {10565} (\bibinfo {year}
  {2016})}\BibitemShut {NoStop}%
\bibitem [{\citenamefont {Simmonds}\ \emph {et~al.}(2011)\citenamefont
  {Simmonds}, \citenamefont {Salter}, \citenamefont {Jesacher},\ and\
  \citenamefont {Booth}}]{Salter2011}%
  \BibitemOpen
  \bibfield  {author} {\bibinfo {author} {\bibfnamefont {R.~D.}\ \bibnamefont
  {Simmonds}}, \bibinfo {author} {\bibfnamefont {P.~S.}\ \bibnamefont
  {Salter}}, \bibinfo {author} {\bibfnamefont {A.}~\bibnamefont {Jesacher}}, \
  and\ \bibinfo {author} {\bibfnamefont {M.~J.}\ \bibnamefont {Booth}},\ }\href
  {http://www.eng.ox.ac.uk/dop/papers/dualAODiamond.pdf} {\bibfield  {journal}
  {\bibinfo  {journal} {Opt. Express}\ }\textbf {\bibinfo {volume} {19}},\
  \bibinfo {pages} {24122} (\bibinfo {year} {2011})}\BibitemShut {NoStop}%
\bibitem [{\citenamefont {Salter}\ and\ \citenamefont
  {Booth}(2012)}]{Salter2012b}%
  \BibitemOpen
  \bibfield  {author} {\bibinfo {author} {\bibfnamefont {P.~S.}\ \bibnamefont
  {Salter}}\ and\ \bibinfo {author} {\bibfnamefont {M.~J.}\ \bibnamefont
  {Booth}},\ }\href@noop {} {\bibfield  {journal} {\bibinfo  {journal} {Opt.
  Express}\ }\textbf {\bibinfo {volume} {20}},\ \bibinfo {pages} {19978}
  (\bibinfo {year} {2012})}\BibitemShut {NoStop}%
\bibitem [{\citenamefont {Burghoff}, \citenamefont {Nolte},\ and\ \citenamefont
  {Tunnermann}(2007)}]{Burghoff2007}%
  \BibitemOpen
  \bibfield  {author} {\bibinfo {author} {\bibfnamefont {J.}~\bibnamefont
  {Burghoff}}, \bibinfo {author} {\bibfnamefont {S.}~\bibnamefont {Nolte}}, \
  and\ \bibinfo {author} {\bibfnamefont {A.}~\bibnamefont {Tunnermann}},\
  }\href@noop {} {\bibfield  {journal} {\bibinfo  {journal} {Appl. Phys. A}\
  }\textbf {\bibinfo {volume} {89}},\ \bibinfo {pages} {127} (\bibinfo {year}
  {2007})}\BibitemShut {NoStop}%
\bibitem [{\citenamefont {Draganski}\ \emph {et~al.}(2012)\citenamefont
  {Draganski}, \citenamefont {Finkman}, \citenamefont {Gibson}, \citenamefont
  {Fairchild}, \citenamefont {Ganesan}, \citenamefont {Nabatova-Gabain},
  \citenamefont {Tomljenovic-Hanic}, \citenamefont {Greentree},\ and\
  \citenamefont {Prawer}}]{draganski2011}%
  \BibitemOpen
  \bibfield  {author} {\bibinfo {author} {\bibfnamefont {M.~A.}\ \bibnamefont
  {Draganski}}, \bibinfo {author} {\bibfnamefont {E.}~\bibnamefont {Finkman}},
  \bibinfo {author} {\bibfnamefont {B.~C.}\ \bibnamefont {Gibson}}, \bibinfo
  {author} {\bibfnamefont {B.~A.}\ \bibnamefont {Fairchild}}, \bibinfo {author}
  {\bibfnamefont {K.}~\bibnamefont {Ganesan}}, \bibinfo {author} {\bibfnamefont
  {N.}~\bibnamefont {Nabatova-Gabain}}, \bibinfo {author} {\bibfnamefont
  {S.}~\bibnamefont {Tomljenovic-Hanic}}, \bibinfo {author} {\bibfnamefont
  {A.~D.}\ \bibnamefont {Greentree}}, \ and\ \bibinfo {author} {\bibfnamefont
  {S.}~\bibnamefont {Prawer}},\ }\href@noop {} {\bibfield  {journal} {\bibinfo
  {journal} {Opt. Mat. Express}\ }\textbf {\bibinfo {volume} {2}},\ \bibinfo
  {pages} {644} (\bibinfo {year} {2012})}\BibitemShut {NoStop}%
\bibitem [{\citenamefont {Sotillo}\ \emph {et~al.}(2016)\citenamefont
  {Sotillo}, \citenamefont {Bharadwaj}, \citenamefont {Hadden}, \citenamefont
  {Sakakura}, \citenamefont {Chiappini}, \citenamefont {Fernandez},
  \citenamefont {Longhi}, \citenamefont {Jedrkiewicz}, \citenamefont
  {Shimotsuma}, \citenamefont {Criante}, \citenamefont {Osellame},
  \citenamefont {Galzerano}, \citenamefont {Ferrari}, \citenamefont {Miura},
  \citenamefont {Ramponi}, \citenamefont {Barclay},\ and\ \citenamefont
  {Eaton}}]{Eaton2016}%
  \BibitemOpen
  \bibfield  {author} {\bibinfo {author} {\bibfnamefont {B.}~\bibnamefont
  {Sotillo}}, \bibinfo {author} {\bibfnamefont {V.}~\bibnamefont {Bharadwaj}},
  \bibinfo {author} {\bibfnamefont {J.~P.}\ \bibnamefont {Hadden}}, \bibinfo
  {author} {\bibfnamefont {M.}~\bibnamefont {Sakakura}}, \bibinfo {author}
  {\bibfnamefont {A.}~\bibnamefont {Chiappini}}, \bibinfo {author}
  {\bibfnamefont {T.~T.}\ \bibnamefont {Fernandez}}, \bibinfo {author}
  {\bibfnamefont {S.}~\bibnamefont {Longhi}}, \bibinfo {author} {\bibfnamefont
  {O.}~\bibnamefont {Jedrkiewicz}}, \bibinfo {author} {\bibfnamefont
  {Y.}~\bibnamefont {Shimotsuma}}, \bibinfo {author} {\bibfnamefont
  {L.}~\bibnamefont {Criante}}, \bibinfo {author} {\bibfnamefont
  {R.}~\bibnamefont {Osellame}}, \bibinfo {author} {\bibfnamefont
  {G.}~\bibnamefont {Galzerano}}, \bibinfo {author} {\bibfnamefont
  {M.}~\bibnamefont {Ferrari}}, \bibinfo {author} {\bibfnamefont
  {K.}~\bibnamefont {Miura}}, \bibinfo {author} {\bibfnamefont
  {R.}~\bibnamefont {Ramponi}}, \bibinfo {author} {\bibfnamefont {P.~E.}\
  \bibnamefont {Barclay}}, \ and\ \bibinfo {author} {\bibfnamefont {S.~M.}\
  \bibnamefont {Eaton}},\ }\href@noop {} {\bibfield  {journal} {\bibinfo
  {journal} {arXiv:1605.01854}\ } (\bibinfo {year} {2016})}\BibitemShut
  {NoStop}%
\end{thebibliography}

The authors gratefully acknowledge the Leverhulme Trust (RPG-2013-044) and the UK Engineering and Physical Sciences Research Council (EP/K034480/1) for financial support.

\end{document}